\documentclass[preprint,12pt]{elsarticle}

\usepackage{graphicx}
\usepackage{amssymb}

\usepackage{url}

\journal{Journal Name}

\begin{document}

\begin{frontmatter}


\title{A deep-learning model for evaluating and predicting the impact of lockdown policies on COVID-19 cases}



\author{Ahmed Ben Said}
\author{Abdelkarim Erradi}
\author{Hussein Ahmed Aly}
\author{Abdelmonem Mohamed}
\address{Computer Science \& Engineering Department, College of Engineering, 2713, Doha, Qatar}
\address{\{abensaid, erradi, ha1601589, am1604044\}@qu.edu.qa}

\begin{abstract}
To reduce the impact of COVID-19 pandemic most countries have implemented several counter-measures to control the virus spread including school and border closing, shutting down public transport and workplace and restrictions on gathering. In this research work, we propose a deep-learning prediction model for evaluating and predicting the impact of various lockdown policies on daily COVID-19 cases. This is achieved by first clustering countries having similar lockdown policies, then training a prediction model based on the daily cases of the countries in each cluster along with the data describing their lockdown policies. Once the model is trained, it can used to evaluate several scenarios associated to lockdown policies and investigate their impact on the predicted COVID cases. Our evaluation experiments, conducted on Qatar as a use case, shows that the proposed approach achieved competitive prediction accuracy. Additionally, our findings highlighted that lifting restrictions particularly on schools and border opening would result in significant increase in the number of cases during the study period.
\end{abstract}


\end{frontmatter}


\section{Introduction}
A new coronavirus (COVID-19) \cite{yang,review}has emerged from Wuhan, the capital of Hubei province in central China in December 2019. The World Health Organization has classified the virus as a global pandemic on March 11 2020  \cite{who}, with more than 280k cases and more than 4000 deaths worldwide.
Several research works and effort have been conducted in order to get better insights about the virus and the factors affecting its fast spread while the world is racing against the clock to develop an effective vaccine. Sun et al. \cite{sun} investigated the correlation between geographic information (latitude, longitude, altitude) and the cumulative infected population under the lockdown policy during the period from December 8, 2019 till April 8, 2020. The authors discovered a negative correlation between the cumulative cases per province and the latitude/longitude. The findings also showed population density is not an important factor in the virus outbreak under strict lockdown measures. Sardar et al. \cite{lockdown_india} assessed the implication of lockdown in some indian states and overall india. A new mathematical model is presented that incorporates lockdown measures and used to estimate new cases from May 17 to 31, 2020. In addition, authors conducted a sensitivity analysis by studying the correlation analysis of two epidemological parameters with the lockdown measures and the basic reproduction number $R_0$. The findings also showed that lockdown is indeed effective in locations with high infection rate. A lift of lockdown restriction after May 17 would result in high spike of number of cases. In the study of Atalan \cite{atalan}, analysis of the effect of the number of lockdown days on the spread of COVID-19 is presented. The study showed evidence that the pandemic can be suppressed by effective lockdown measures. Thee author argue also that these measures are also effective from psychological, environmental and economical perspectives. Vinceti et al. \cite{italy} analyzed mobility restriction  in the three most affected Italian regions, Lombardy, Veneto and Emilia-Romagna, from February 1 through April 6, 2020. Results showed that the daily number of cases is inversely correlated to mobility restriction after the second lockdown policy. The peak is witnessed 14 to 18 days from imposing the lockdown. A study of the epidemiological situation in the french region of  \^{I}le de France is presented in \cite{france}. The authors a stochastic age-structured transmission model that includes age profiles and social contacts in the area of study. This model is used to evaluate the lockdown policy. The simulated scenarios are integrated by introducing changes to the contact matrices. The lockdown information is derived from mobility data provided by mobile phone. Simulation scenarios are established includig testing with different types and duration of social distancing. Results showed that the prior to lockdown, the estimated reproduction number is 3.18. It falls to 0.68 during lockdown thanks to 81\% reduction in number of contacts.\newline
Although great research efforts have been dedicated to investigate the impact of restriction measures on the spread of the pandemic, to the best of our knowledge, no reported work has investigate the impact of multiple restrictions measures e.g. school closing, restriction on gathering, workplace closing, travel restriction, public transport shutdown, and how any change in the restriction would affect the pandemic situation. In this work, we proposed a deep learning based approach that enables stakeholders and decision makers to establish several scenarios related to restriction measures and investigate their impact. In addition, the proposed solution is versatile, in the sense that it is applicable for any country. In addition, the proposed method considers  countries sharing similar restrictions and lockdown policies to analyze establish the simulated scenarios and assess their implications, rather than analyzing each country separately
\section{What If analysis}
Our objective is to propose a solution to assist decision makers to establish multiple scenarios related to restrictions imposed to control the spread of COVID-19. In the followings, we present an overview of the proposed approach and its technical details.
\subsection{What if framework}
\begin{figure}[h!]
	\centering
	\includegraphics[scale=.8]{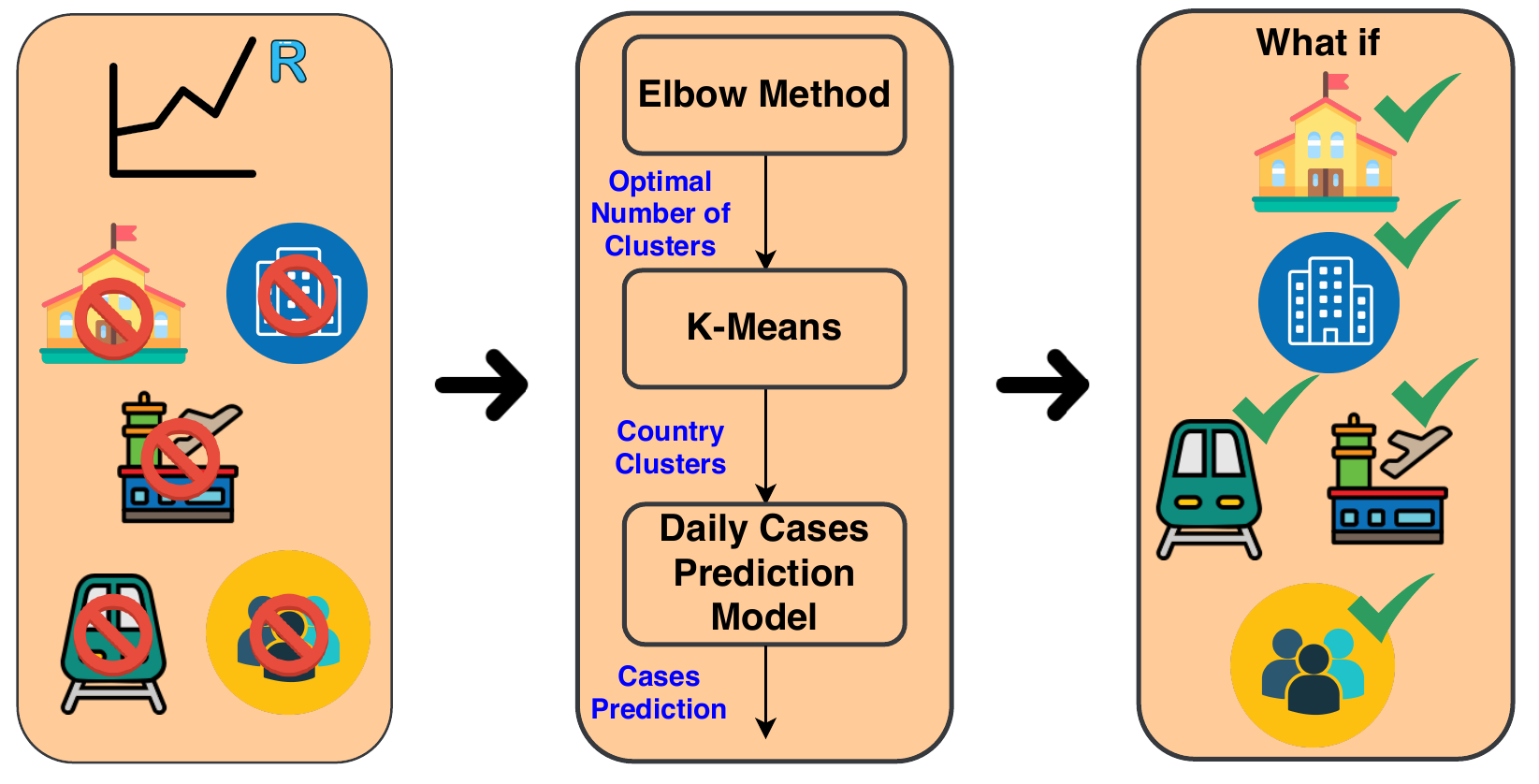}
	\caption{What If analysis framework}
	\label{framework}
\end{figure}
Fig. \ref{framework} illustrates the proposed framework. First, we analyze the lockdown strategy of each country in terms of:
\begin{itemize}
    \item How fast the country reacted to the spread of the virus: this is assessed in terms of difference in days between the first case is reported and the day the following restrictions are imposed: school closing, restriction on gatherings, border closing, public transport shutdown and workplace closing.
    \item Efficiency of the lockdown policy: we propose to assess the efficiency in terms of reproduction number. This specific number is the main concern of all health authorities and decision makers. Indeed, a value greater than 1 indicates that the virus is spreading while a value less than 1 indicates that the virus spread is being controlled. We propose to calculate this number for reach country and averaged every two weeks.
\end{itemize}
This set of extracted features are intended to be clustered in order to determine countries sharing similar lockdown and restriction policies. For this, we use the well-known K-Means algorithm which requires the number of clusters as input paramter. Hence, we use Elbow method to determine the optimal number of cluster. For a given country we intend to analyze, we use its daily reported cases as well as to the daily reported cases of the countries belonging to the same cluster. These data combined with the lockdown measures are used to train a deep learning model to predict the next day number of COVID-19 cases. Once trained, this learning model is queried with data reflecting a lockdown scneario. In other words, we are "asking" the model to predict the next day number of cases if, for example, schoolo are open and/or restriction on gathering is restricted while other measures remain effective.
\newline
In the following, we present the technical details of the reproduction number calculation, Elbow method, K-Means clustering and the deep learning-based prediction model.
\subsection{Reproduction number}
Bettencourt et al. \cite{rt} proposed a Bayesian approach to estimate the reproduction number in real time $R_t$. This value depends on the value of the previous day $R_{t-1}$ and the all the previous $m$ values $R_{t-m}$. Using Bayesian modelling, the belief about the true value $R_t$ is updated based on how many reported cases each day:
\begin{equation}
    P(R_t|k) = \frac{P(k_t|R_t)P(R_t)}{P(k_t)}
\end{equation}
where $P(R_t)$ is the prior belief about $R_t$ without data, $P(k|R_t)$ is the conditional probability of having $k$ cases given $R_t$ and $P(k)$ is the probability of having $k$ cases. By using the posterior of the previous day $P(R_{t-1}|k_{t-1})$ as today prior $P(R_t)$, then we can write
\begin{equation}
    P(R_t|k_t) \propto P(k_t|R_t)P(R_{t-1}|k_{t-1})
\end{equation}
By iterating across all periods, we obtain:
\begin{equation}
    P(R_t|k_t) \propto P(R_0) \prod_{t=0}^T P(k_t|R_t)
\end{equation}
Assuming a uniform prior $P(R_0)$, we have:
\begin{equation}
    P(R_t|k_t) \propto \prod_{t=0}^T P(k_t|R_t)
\end{equation}
However, as emphasized by \cite{rt}, for $R_t$ that remains greater than 1 for a long period and then becomes less than 1, the posterior gets stuck. In other words, the posterior cannot forget about long period of times for which $R_t>1$. In his study of reproduction number for United States, Kevin Systrom \cite{kevin} suggested including into the posterior only the previous $m$ days. Hence, the prior is derived using information from the recent past rather than the entire history. Thus, we can write:
\begin{equation}
     P(R_t|k_t) \propto \prod_{t=T-m}^T P(k_t|R_t)
\end{equation}
\subsection{Identifying countries with common lockdown policy}
The objective to group and identify countries having similar lockdown policy. The intuitive idea is that COVID-19 would impact these countries the same way. To achieve this goal, we apply the Elbow method to determine the optimal number of clusters. This optimal number is used as input parameter for the K-Means clustering algorithm. In the following, we present the technical details of each step.
\subsubsection{Elbow method}
Let $X=\{x_1,x_2,...,x_n\}$ be $n$ of d-dimensional points to be clustered into $K$ clusters, i.e. assigning each $x_i$, $i=1,...,n$ to a cluster $c_k$, $k=1,...,K$. K-Means partitions the data by minimizing the squared error between the mean of a cluster and the data points, members of the clusters. Let $m_k$ be the mean of cluster $c_k$. The squared error between a cluster center and its members is defined as:
\begin{equation}
    J(c_k) = \sum\limits_{x_i \in c_k} || x_i - m_k||^2 
\end{equation}
K-Means seeks to minimize the sum of the squared errors:
\begin{equation}
    J(C) = \sum\limits_{k=1}^K \sum\limits_{x_i \in c_k} || x_i - m_k||^2
    \label{kmeans}
\end{equation}
Where $C$ is the set of clusters. To minimize Eq. \ref{kmeans}, the following steps are applied:
\begin{enumerate}
    \item Randomly assign $K$ cluster centers and repeat step 2 and 3.
    \item Assign each data point to the closest cluster center.
    \item Calculate the new cluster centers.
\end{enumerate}
The number of cluster $K$ is an input parameter for K-Means. Hence, we use the Elbow method to determine the optimal number of clusters for which the obtained partition is compact, i.e. low $J(C)$. By adding more clusters would result in even more compact partition which leads to over-fitting. Hence, the variation of $J(C)$ with respect to $K$ would exhibit first a fast decrease followed by a slow one. The Elbow method recommends to select the number of cluster that corresponds to the elbow of the curve $J(C)$ vs $K$.
\subsection{Prediction model}
To conduct the 'What if' analysis for a specific country, the cluster to which the contry belongs is idenfied. Then, the daily COVID-19 cases of these countries are collected in addition to data related to lockdown measures. In the followings, we detail characteristics of  data and prediction model trained on both daily COVID-19 and lockdown data.
\subsubsection{Data description}
The temporal characteristic is a critical component of the data as data are collected on daily basis. Hence, it can be seen as multivariate time series and consists of:
\begin{itemize}
    \item Daily COVID-19 cases. These data are  provided by government agencies It can be collected through several APIs for this information. We collect data from February 15th to July 31st.
    \item School closing, where 0 indicates no measures are taken, 1- recommend closing, 2- require closing (only some levels or categories, e.g. just high school, or just public schools) and
    3- require closing all levels.
    \item Workplace closing, where 0 indicates no measures are taken, 1- recommend closing (or recommend work from home), 2- require closing (or work from home) for some sectors or categories of workers and 3- require closing (or work from home) for all-but-essential workplaces (e.g. grocery stores, doctors).
    \item Restrictions on gatherings: where 0 indicates no restrictions are imposed, 1- restrictions on very large gatherings (the limit is above 1000 people), 2- restrictions on gatherings between 101-1000 people, 3- restrictions on gatherings between 11-100 people, 4- restrictions on gatherings of 10 people or less.
    \item Public transport shutdown where 0 indicates no measures are taken, 1- recommend closing (or significantly reduce volume/route/means of transport available) and 2- require closing (or prohibit most citizens from using it)
    \item International travel controls where 0 indicates no restrictions are taken,1- screening arrivals, 2- quarantine arrivals from some or all regions, 3- ban arrivals from some regions and 4- ban on all regions or total border closure.
\end{itemize}
\subsubsection{Two-pathway model}
\begin{figure}[h!]
	\centering
	\includegraphics[scale=.75]{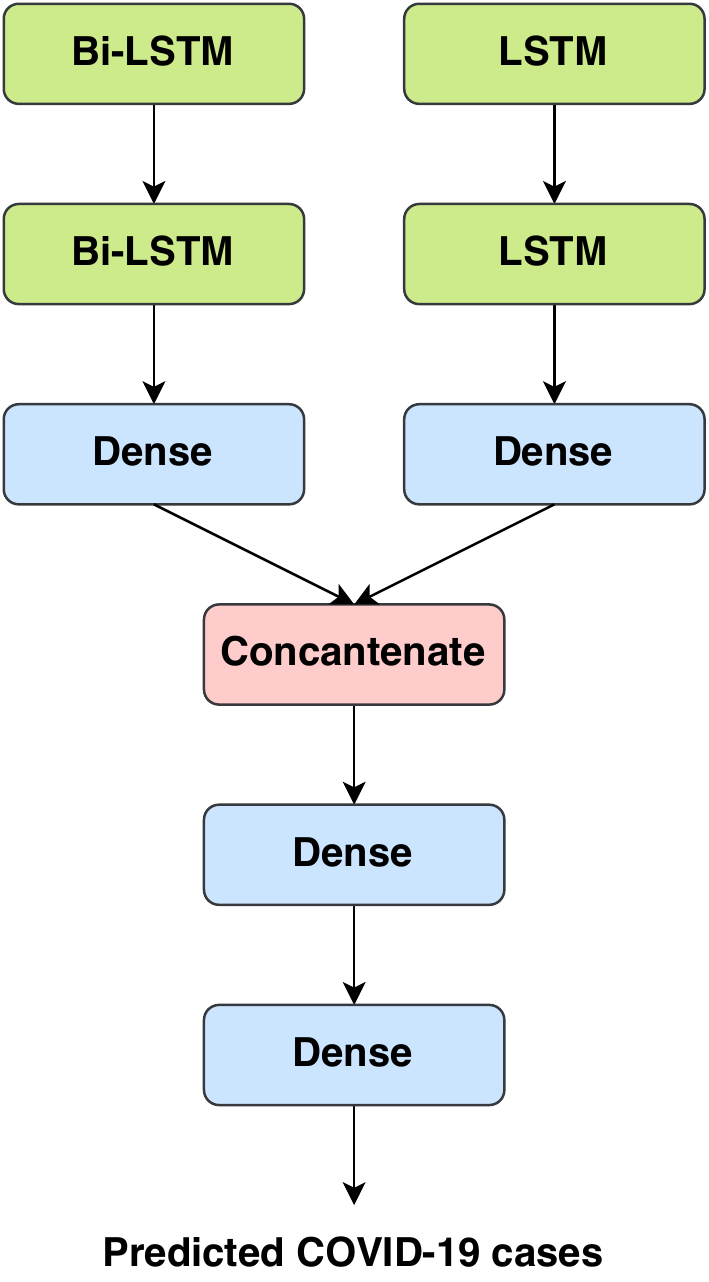}
	\caption{Two-pathway model}
	\label{model}
\end{figure}

Fig. \ref{model} depicts the general architecture of the proposed daily COVID-19 forecasting model. It is characterized by two pathways: one pathway dedicated to daily COVID-19 time series data while the other pathway is fed with multivariate time series of lockdown measures. The first pathway is consists of a stack of two Bidirectional LSTM layers followed by a dense layer. The second pathway is a stack of two LSTM layers followed by a dense layer. On top of these pathways, a merge layer is added where concatenation of the two dense layers is performed. The merge layer is followed by two dense layers. Th model outputs the prediction of the next day number of cases. 
\newline
In the followings, we present the technical details of the LSTM and Bidirectional LSTM layers.
\begin{itemize}
    \item\textbf{LSTM layer:}
    An LSTM layer consists of a sequence of LSTM cells and the sequence data are fed in a forward way. The LSTM cell depicted in Fig. \ref{lstm_cell}.
\begin{figure}[h!]
	\centering
	\includegraphics[scale=.7]{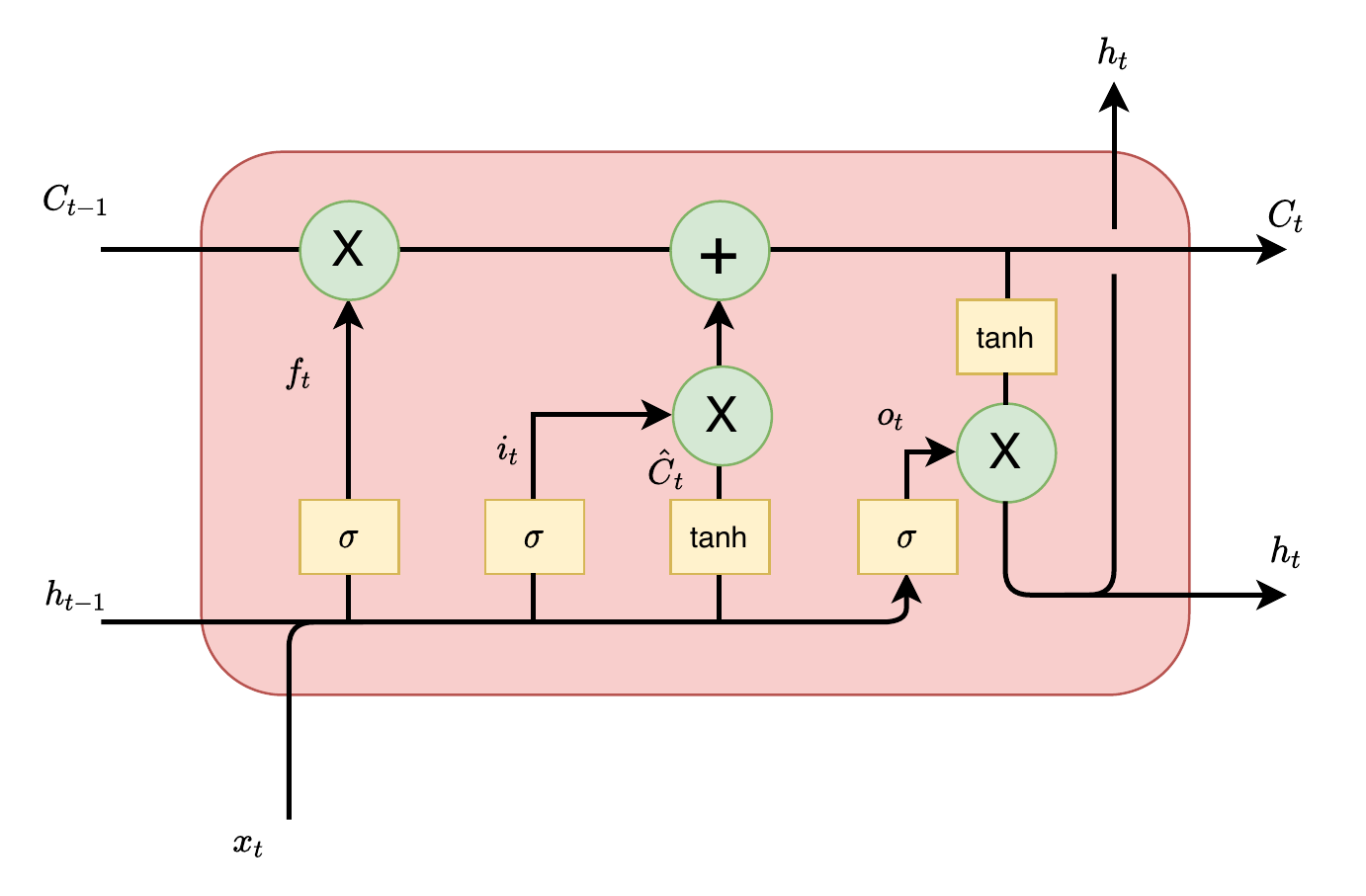}
	\caption{Long Short-Term Memory (LSTM) cell}
	\label{lstm_cell}
\end{figure}
Given the current value $x_t$, the previous hidden state $h_{t-1}$ and the previous state $C_{t-1}$, the following transformations are applied:
\begin{equation}
    f_t = \sigma\Big(W_f\cdot [h_{t-1}, x_t] + b_f \Big)
\end{equation}
\begin{equation}
    i_t=\sigma\Big(W_i[h_{t-1},x_t] + b_i\Big)
\end{equation}
\begin{equation}
    \hat{C}_t=tanh\Big(W_C[h_{t-1},x_t] +b_c\Big)
\end{equation}
\begin{equation}
    C_t=f_t*C_{t-1}+i_t*\hat{C}_t
\end{equation}
\begin{equation}
    o_t = \sigma\Big( W_o[h_{t-1},x_t]+b_o\Big)
\end{equation}
\begin{equation}
    h_t = o_t*tanh(C_t)
\end{equation}
Where $\sigma$ and $tanh$ are the sigmoid and hyperbolic tangent function respectively. $f_t$ is the forget gate, $i_t$ is the input gate and $o_t$ is the output gate. $W$ and $b$ are the weight matrix and bias vector respectively. $[\cdot]$ is the concatenation operator and $*$ is the dot product.
\item \textbf{Bidirectional LSTM (Bi-LSTM):}  Bi-LSTM includes another LSTM layer for which the data is fed in backward way as depicted in Fig. \ref{bi-lstm}.
\begin{figure}[h!]
	\centering
	\includegraphics[scale=.75]{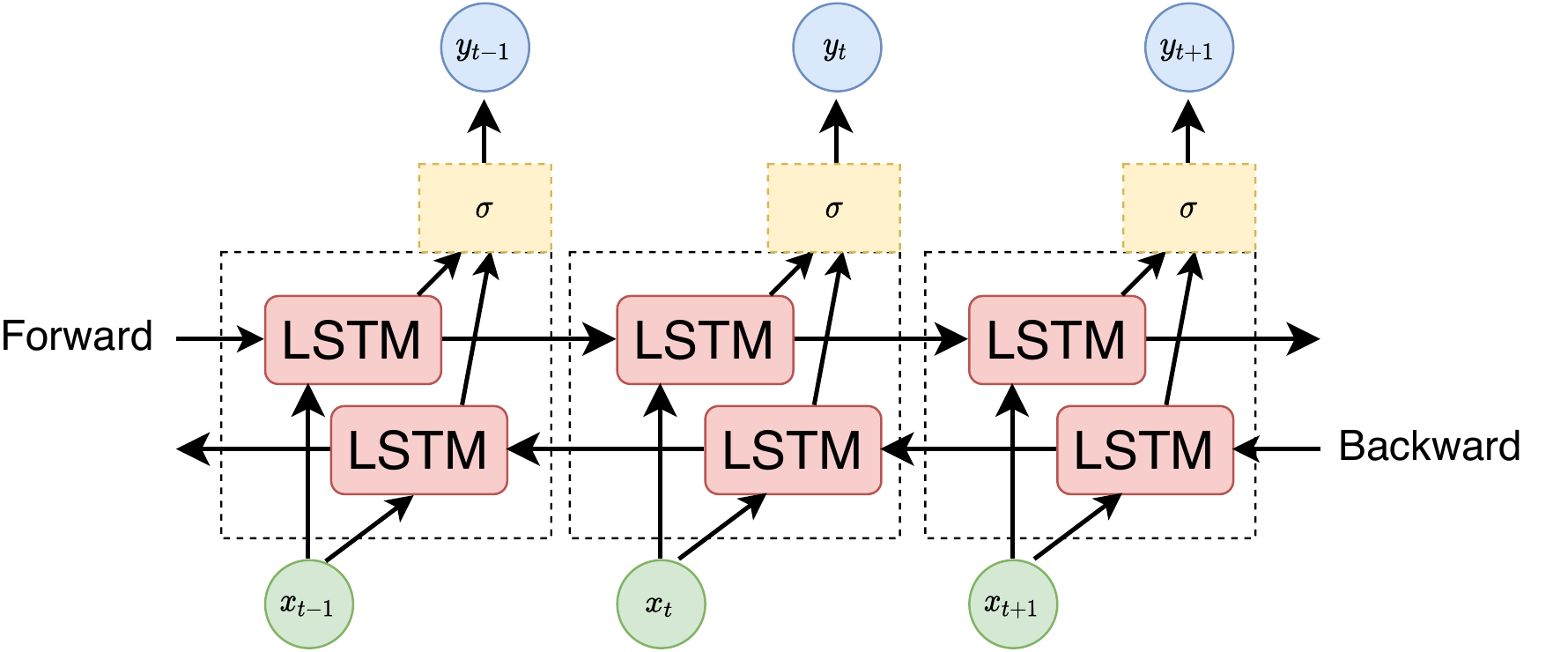}
	\caption{Bidirectional Long Short-Term Memory (LSTM)}
	\label{bi-lstm}
\end{figure}
\end{itemize}

\section{Experimental results}
In this section, we focus on Qatar as a use-case. First, we assess the proposed method by assessing the prediction accuracy, i.e. the error between the actual and predicted daily COVID-19 cases in Qatar. We compare our proposed solution with two other approaches: same model trained without lockdown data and model trained on data of Qatar daily COVID-19 cases only. Then, we investigate the model outcome while changing several input parameters associated to the lockdown measures.
\subsection{prediction of daily COVID-19 cases}
\begin{figure}[h!]
	\centering
	\includegraphics[scale=.6]{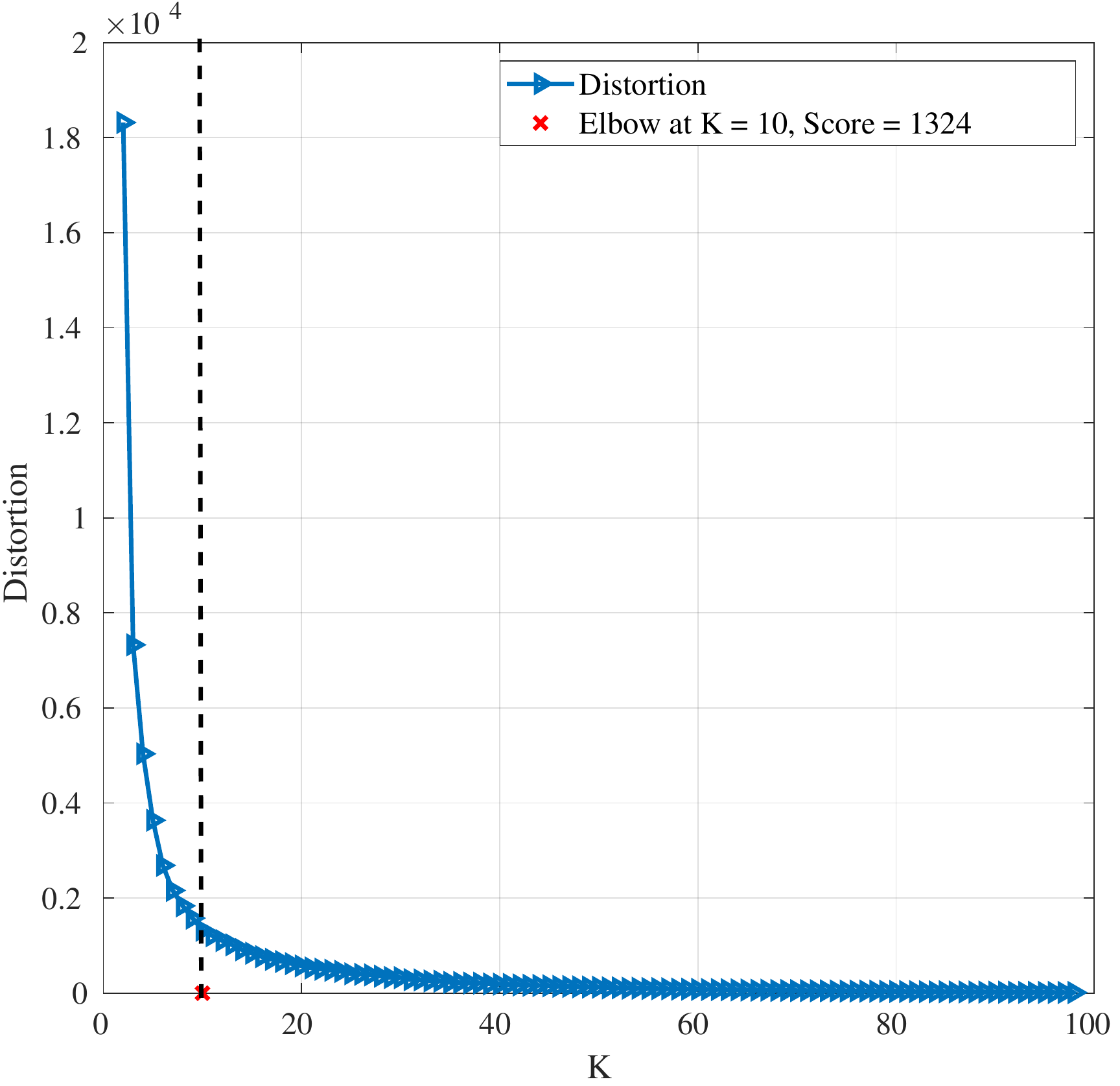}
	\caption{Distortion score for different numbers of clusters.  Elbow corresponds to K= 10}
	\label{elbow}
\end{figure}

\begin{figure}[h!]
	\centering
	\includegraphics[scale=.6]{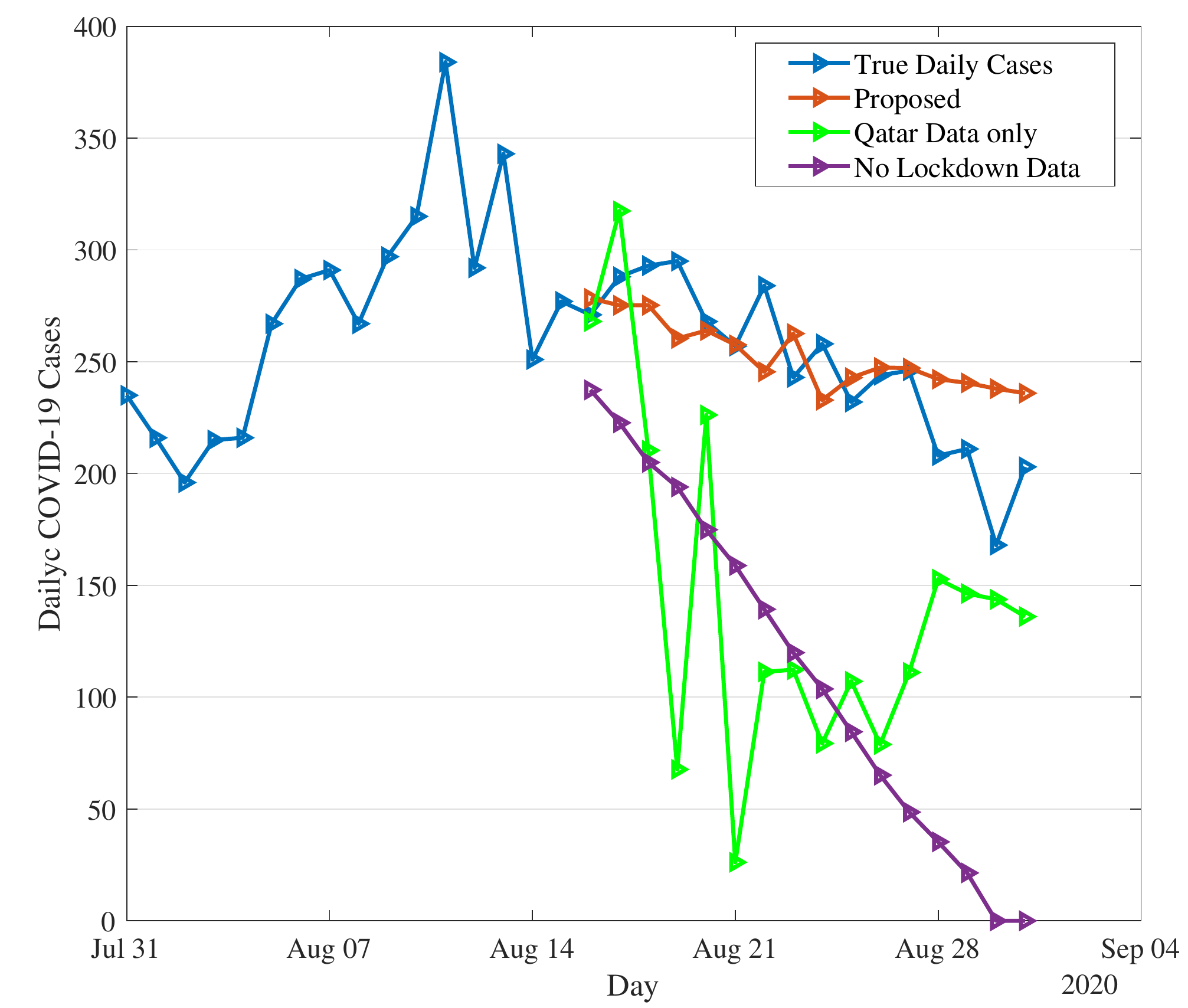}
	\caption{Distortion score for different numbers of clusters.  Elbow corresponds to K= 10}
	\label{prediction}
\end{figure}

Fig. \ref{elbow} illustrated the Elbow method results. Results show that the optimal number of clusters correspond to $K=10$ with a distortion score = 1324.
This number is used as input parameter to K-Means algorithm. Clustering results shows that Qatar belongs to a cluster with other countries including Azerbaijan, Benin, Bahrain, Georgia, Croatia, Indonesia, Italy, Kuwait, Lebanon, Mexico, Mozambique, Norway, Oman, Pakistan, Romania, Seychelles. These countries share similar characteristics in terms of how fast they reacted to the first reported cases and the evolution of the reproduction number $R_t$. They are characterized by an average $Rt=1.1$ meaning that the the virus spread is almost contained.
\newline
Fig. \ref{prediction} shows Qatar daily cases during the month of August in addition to three model predicted cases starting from August 15th. These findings prove the improtance of including lockdown information to predict the daily cases. Indeed, when trained without lockdown data, the model predict continuous decrease of number of cases. When trained on Qatar data only, the model exhibits sharp fluctuation of number of cases. The proposed approach achieves the best prediction performance and able to provide prediction which is close to the actual daily cases.
Table \ref{tab_comparison} details results of quantitative evaluation of the prediction accuracy by calculating the Root Mean Square Error (RMSE) and Mean Absolute Error (MAE of the three models. Results confirm that the proposed approach achieves significant better performance with the lowest RMSE and MAE.
\begin{table}[h!]
\centering
\begin{tabular}{|c|c|c|}
\hline
Approach & RMSE & MAE \\ \hline
Proposed & 28 & 21 \\ \hline
Qatar Data only & 129 & 108 \\ \hline
No Lockdown Data & 197 & 161 \\ \hline
\end{tabular}
\label{tab_comparison}
\end{table}
\subsection{Effect of lockdown policy changes on daily COVID-19 cases}
\begin{figure}[h!]
	\centering
	\includegraphics[scale=.4]{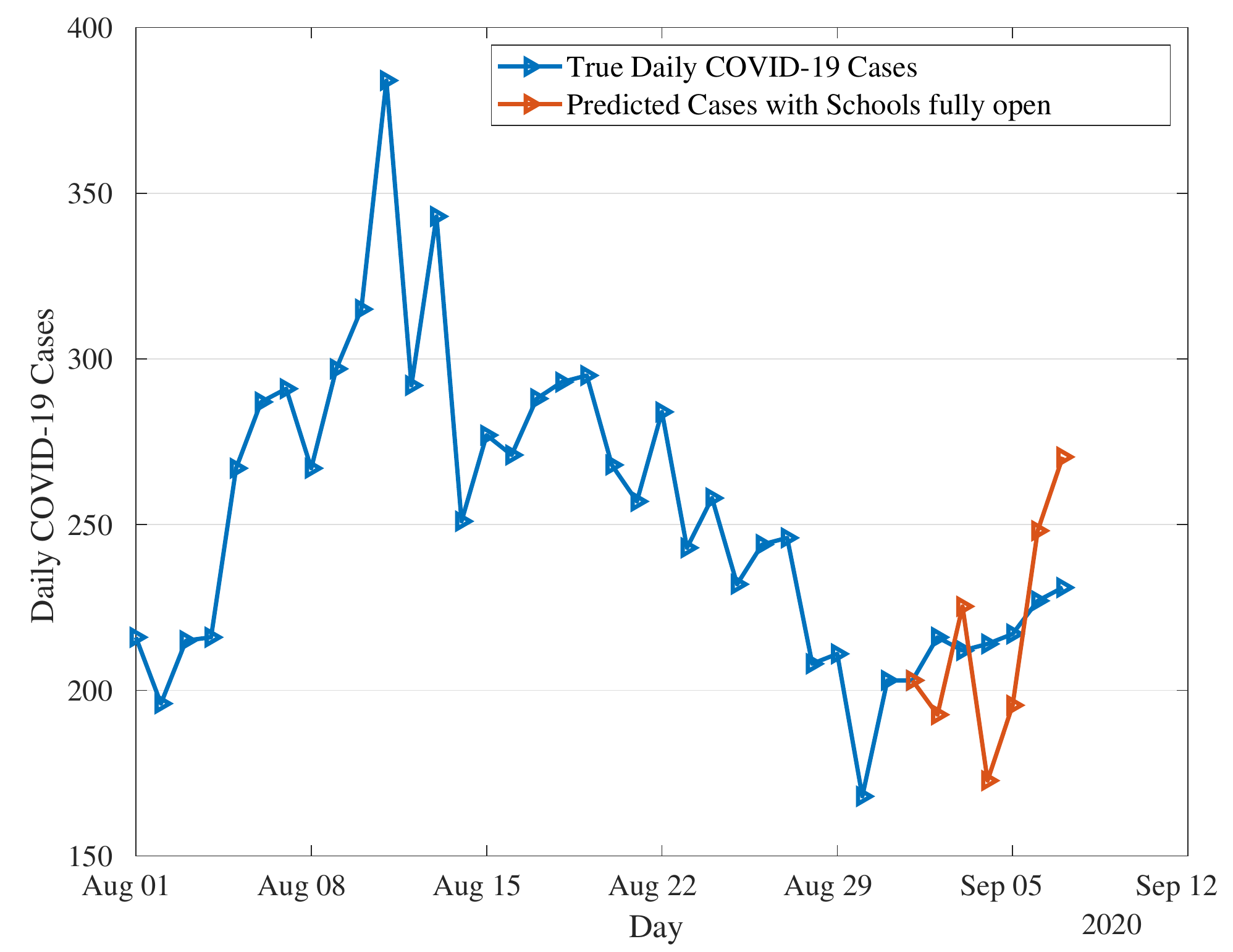}
	\caption{Effect of lifting all restrictions on school in Qatar}
	\label{school}
\end{figure}

\begin{figure}[h!]
	\centering
	\includegraphics[scale=.45]{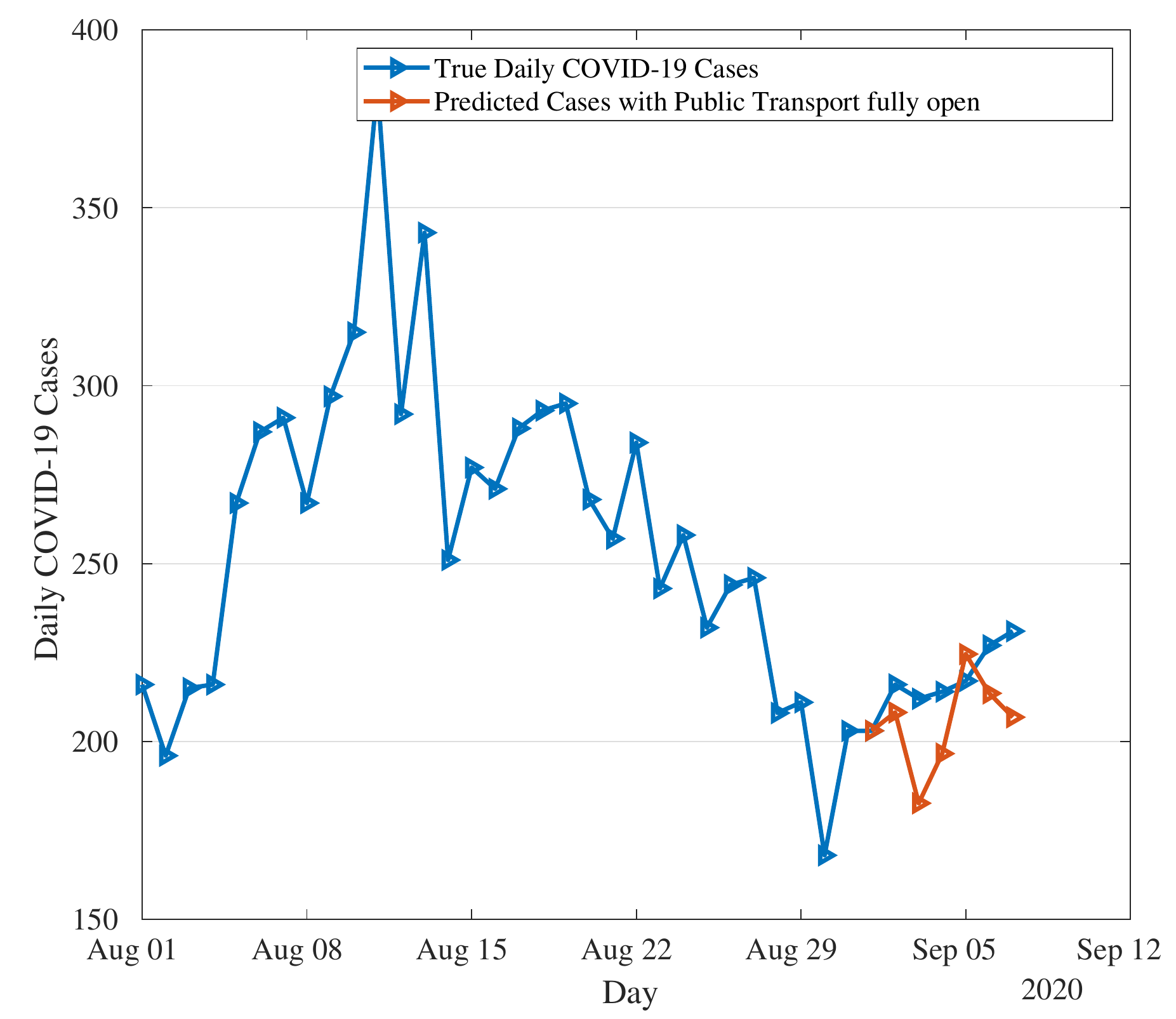}
	\caption{Effect of lifting all restrictions on public transport in Qatar}
	\label{transport}
\end{figure}

\begin{figure}[h!]
	\centering
	\includegraphics[scale=.45]{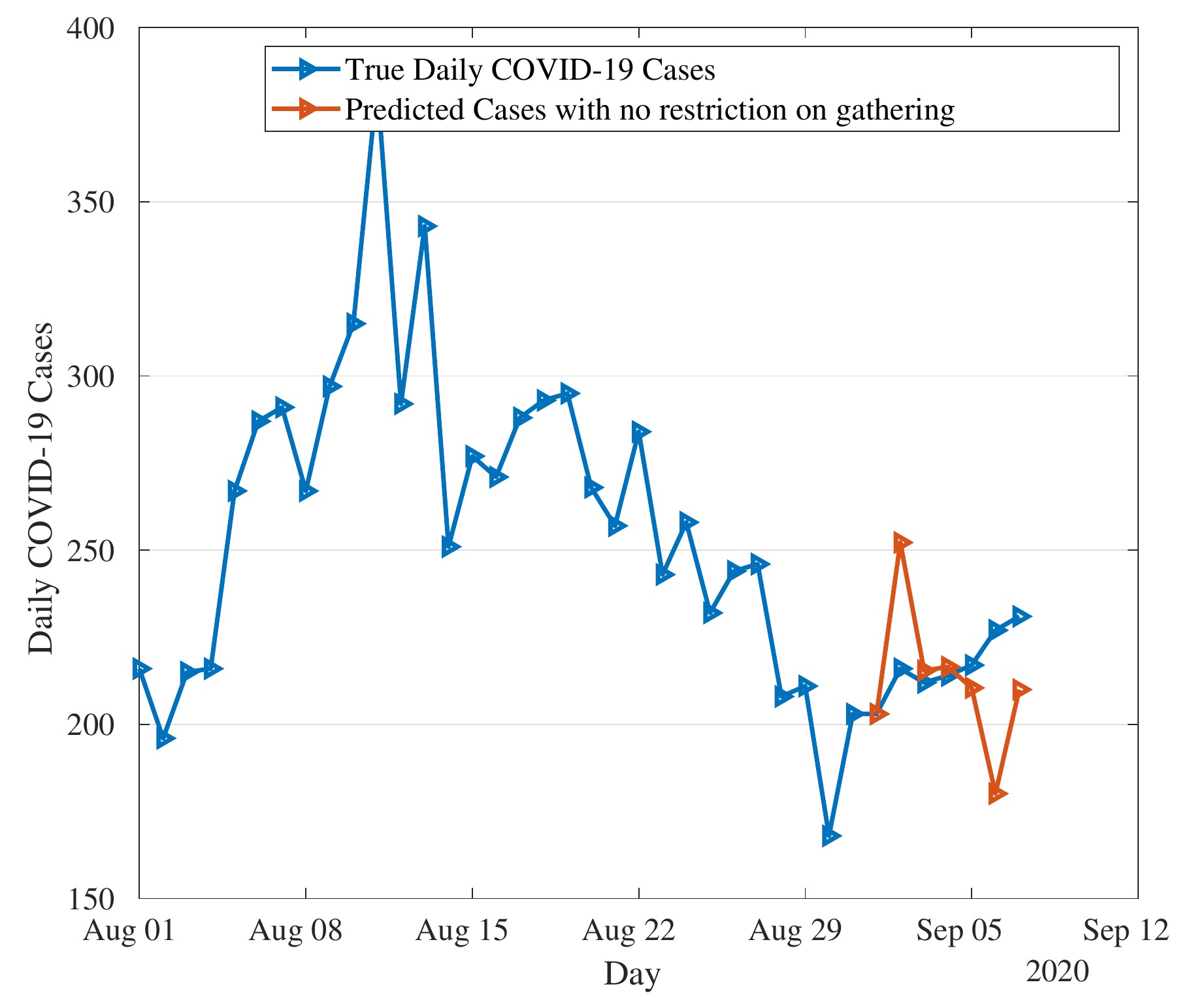}
	\caption{Effect of lifting all restrictions on gathering in Qatar}
	\label{gathering}
\end{figure}

\begin{figure}[h!]
	\centering
	\includegraphics[scale=.45]{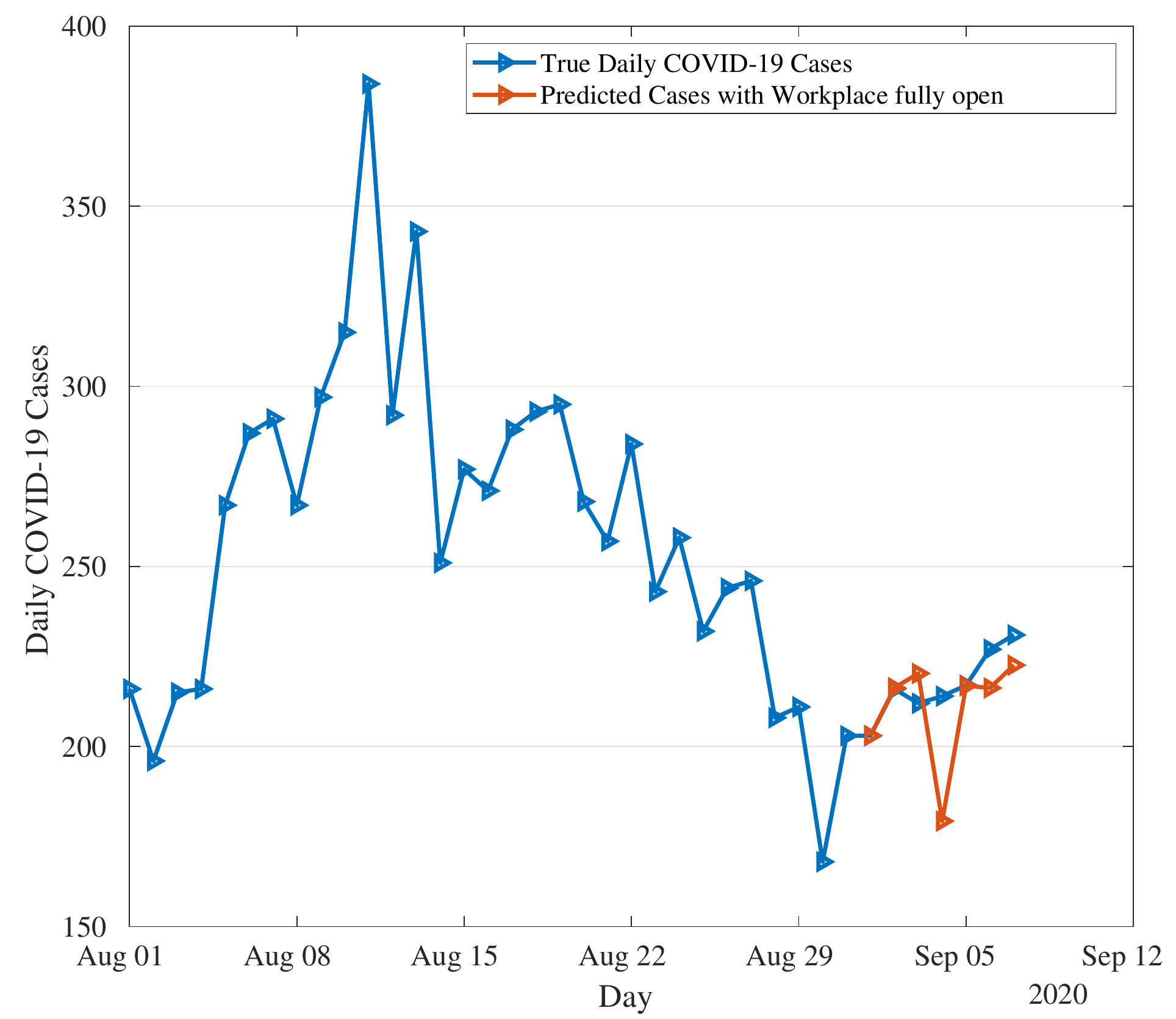}
	\caption{Effect of lifting all restrictions on workplace in Qatar}
	\label{workplace}
\end{figure}

\begin{figure}[h!]
	\centering
	\includegraphics[scale=.48]{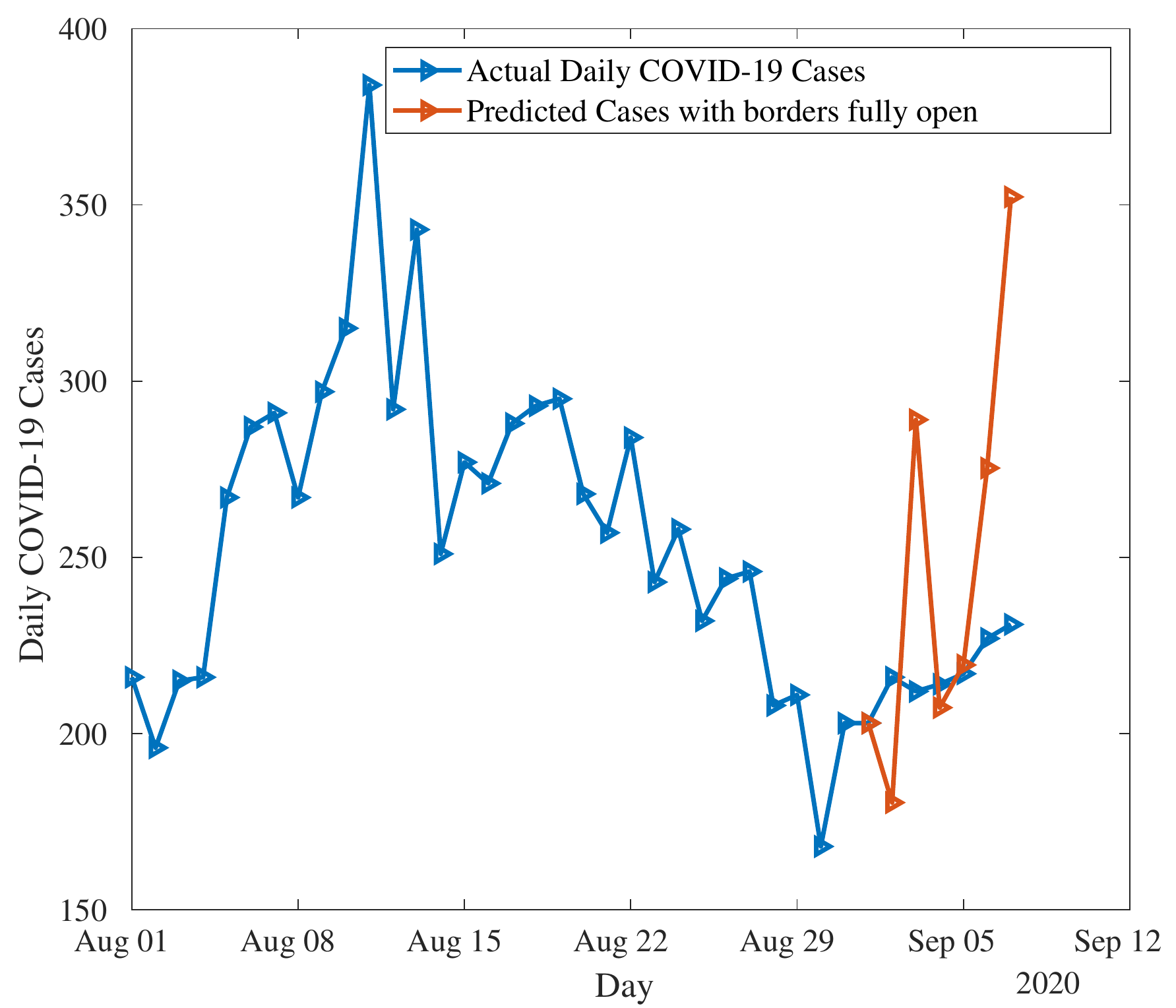}
	\caption{Effect of lifting all restrictions on borders in Qatar}
	\label{borders}
\end{figure}

In this section, we use the proposed model as a baseline to assess Qatar lockdown policy during the first week of September. The proposed model 
can be used to investigate the effect of changing the lockdown policy on the evolution of COVID-19 cases. On September 1st, Qatar entered the 4th phase of its lockdown policy which consists of: partially lifting restrictions on gathering, partially opening public transport and educational institutions. Workplace are opened with 80\% capacity. Travelers entering Qatar from low risk countries are still required to self-quarantine at home for one week. For other countries, passengers must be quarantined at hotels for 15 days. In this section, we analyze how lifting all restrictions on  school, public transport and workplace and border opening would impact the number of cases in Qatar during the first week of September 2020. We establish multiple scenarios in which we hypothetically completely lift restriction on one sector while keeping all other sectors under their actual policy and investigate the model outcome.
\newline
Fig. \ref{school} illustrates how lifting all restrictions on school and educational institutions would impact the daily cases in Qatar starting from 1st of September. Results show that the proposed model predicts a fluctuating numbers till 3rd of September then a continuous alarming increase is witness starting from 4th of September. The actual number of cases is almost stable. During the period of this analysis, school and educational institution policy mandates a partial opening. Indeed, at Qatar University for example, most of the lectures are conducted online for this period.
Fig. \ref{transport} shows the evolution of number of cases if all restrictions on public transport are lifted. We notice that the proposed model predicted a fluctuating number of cases for the one week analysis period without any significant impact. 
We illustrate in Fig. \ref{gathering} the effect of lifting all restrictions on gathering. No specific pattern is detected which suggests that fully opening public transport service would not significantly affect the daily cases.
Fig. \ref{workplace} depicts the predicted number of cases if restriction on workplace is completely lifted. We notice fluctuating number of cases close to the actual numbers which suggest that lifting restriction on workplace while keeping all other restrictions would not affect COVID-19 cases. 
Effect of lifting all restrictions on borders is detailed in Fig. \ref{borders}. Results show dramatic implication of such decision. Indeed, an exponential growth of number of cases would occur with more than 350 cases expected with difference of more than 100 cases compared to the actual number. During the period of this analysis, passengers countries identified as low risk are required to quarantine for one week while passengers from other countries must quarantine at hotels for two weeks.
Our analysis has identified school, educational institution and border restrictions as key factors affecting the daily COVID-19 cases. When restrictions are fully lifted, the proposed model predicted a sudden increase in number of cases indicating that taking such decisions would lead to dramatic consequences. The propose model did not detect any significant changes in number of cases if restriction is fully lifted on gathering, workplace and public transport.
\section{Conclusion}
We proposed a data driven approach aiming at predicting the daily COVID-19 cases which allows also testing several scenarios related to lockdown policy. The proposed model considered both lockdown information and daily cases of countries having similar lockdown policy and showed same response to the outbreak of the virus. We focused in our experiments on Qatar as a use case and showed that the proposed model achieved better prediction by including lockdown information and training model on data of countries with similar policies. Our analysis also showed that completely lifting restrictions on schools and borders would contribute to sudden increase of number of cases in Qatar.  

\section*{Acknowledgment}
This work was made possible by COVID-19 Rapid Response Call (RRC) grant \# RRC-2-104 from the Qatar National Research Fund (a member of Qatar Foundation). The statements made herein are solely the responsibility of the authors.

\bibliographystyle{elsarticle-num-names}

\end{document}